\documentclass[12pt]{iopart}
\usepackage{makeidx,enumerate,graphicx,graphics,afterpage,cite,setstack,amssymb,float,url}

\begin{document}

\title[FPGA-TDC in muon radiography]{Implementation of sub-nanoseconds TDC in FPGA: applications to time-of-flight analysis in muon radiography}

\author{Jacques Marteau$^1$, Jean de Bremond d'Ars$^2$, Dominique Gibert$^{2,3}$, Kevin Jourde$^3$, and Serge Gardien$^1$, Claude Girerd$^1$, Jean-Christophe Ianigro$^1$}

\address{$^1$ Institut de Physique Nucl\'eaire de Lyon, Univ Claude Bernard, UMR 5822 CNRS, Lyon, France.}
\address{$^2$ G\'eosciences Rennes, Univ Rennes 1, UMR 6118 CNRS, Rennes, France.}
\address{$^3$ Institut de Physique du Globe de Paris, Sorbonne Paris Cit\'e, Univ Paris Diderot, UMR 7154 CNRS, Paris, France.}

\begin{abstract}
Time-of-flight ({\it tof}) techniques are standard techniques in high energy physics to determine particles propagation directions. Since particles velocities are generally close to $c$, the speed of light, and detectors typical dimensions at the meter level, the state-of-the-art tof techniques should reach sub-nanosecond timing resolution. Among the various techniques already available, the recently developed ring oscillator TDC ones, implemented in low cost FPGA, feature a very interesting figure of merit since a very good timing performance may be achieved with limited processing ressources. This issue is relevant for applications where unmanned sensors should have the lowest possible power consumption. Actually this article describes in details the application of this kind of {\it tof} technique to muon tomography of geological bodies. Muon tomography aims at measuring density variations and absolute densities through the detection of atmospheric muons flux's attenuation, due to the presence of matter. When the measured fluxes become very low, an identified source of noise comes from backwards propagating particles hitting the detector in a direction pointing to the geological body. The separation between through-going and backward-going particles, on the basis of the {\it tof} information is therefore a key parameter for the tomography analysis and subsequent previsions.
\end{abstract}

\pacs{1315, 9440T}

\maketitle

\section{Introduction}

Density radiography with atmospheric muons aims at determining the density variations or the absolute densities of geological or large volume bodies. The density is measured through the screening effect on the incident muons flux induced by the presence of matter, like for the X rays in a standard medical radiography. The physics of muons tomography may be found elsewhere (Alvarez 1970; Nagamine 1995; Nagamine et al. 1995; Tanaka et al. 2001; Lesparre et al. 2010). For a detailed presentation of recent applications of this muon tomography technique, please refer to Jourde et al. (2013), and references therein.

We will focus in this article on applications in volcanology, where the maximal dimension of the volcano to probe is typically at the kilometer scale. This is for instance the case of La Soufri\`ere de Guadeloupe in the Lesser Antilles or the top craters of Mount Etna in Sicily, Italy (Lesparre et al. 2010; Carbone et al. 2013, Jourde et al. 2013). In both cases, the detectors used in the measurements are set on the slopes of the volcano and their orientation is such that they may catch muons incident from the ``open'' sky, ie not crossing the volcano, and muons which propagated through the volcano. Given the position of the detector and the harsh environmental conditions, the instrument may be designed as robust as possible and totally autonomous (remote control, low power consumption compatible with solar panels or wind wings power sources). Despite these constrains the figure of merit of those detectors is the signal-to-noise ratio to reach the best possible sensitivity. In particular recent applications aimed at detecting small-density constrasts, which translates in small variations in the muons flux to be detected, and put severe constraints on the sensitivity (Nagamine 2003; Lesparre et al. 2010).

\begin{figure}[htbp]
\centerline{\includegraphics[height=6cm,width=0.5\linewidth]{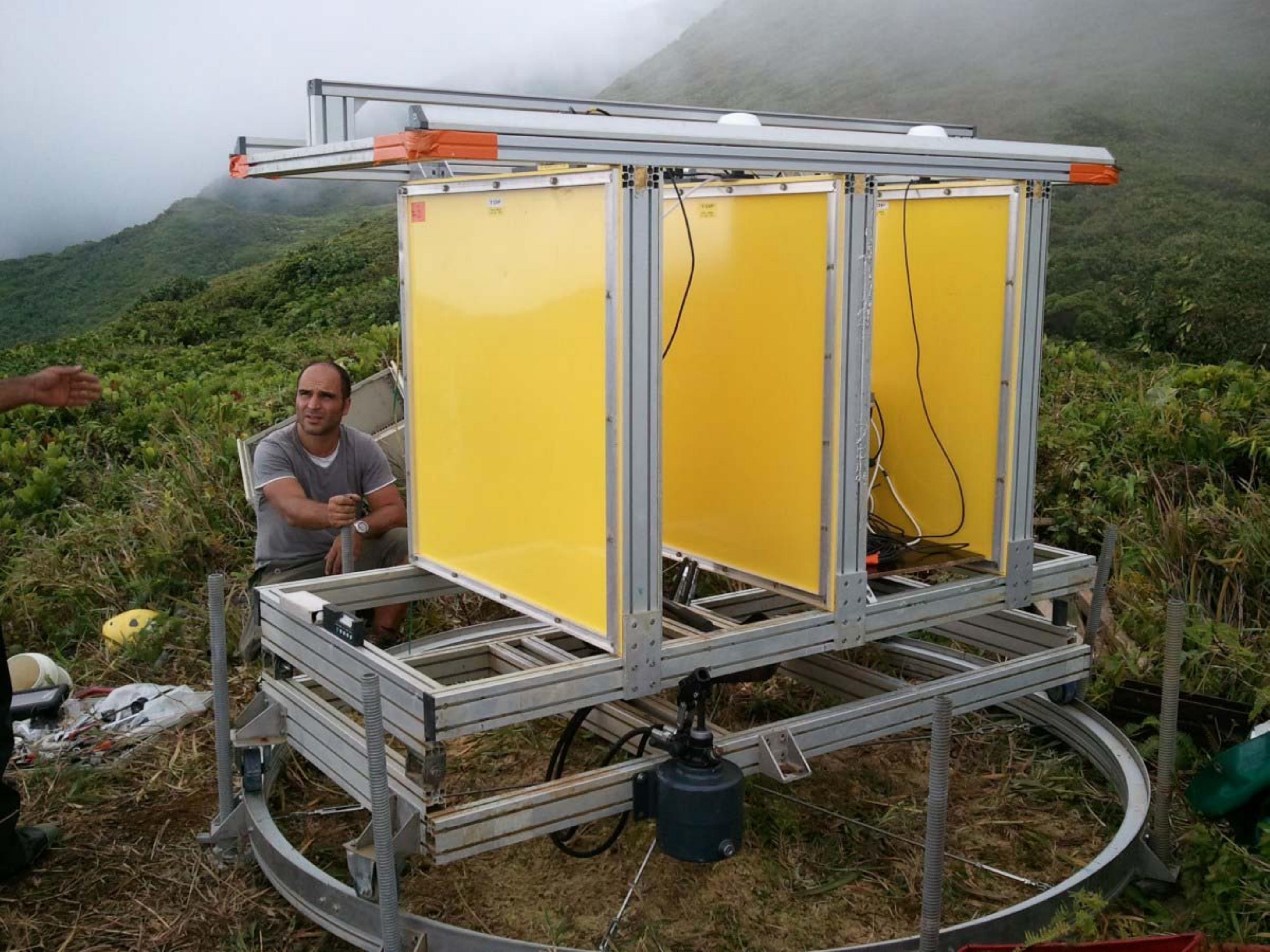}\hfill
\includegraphics[height=6cm,width=0.4\linewidth]{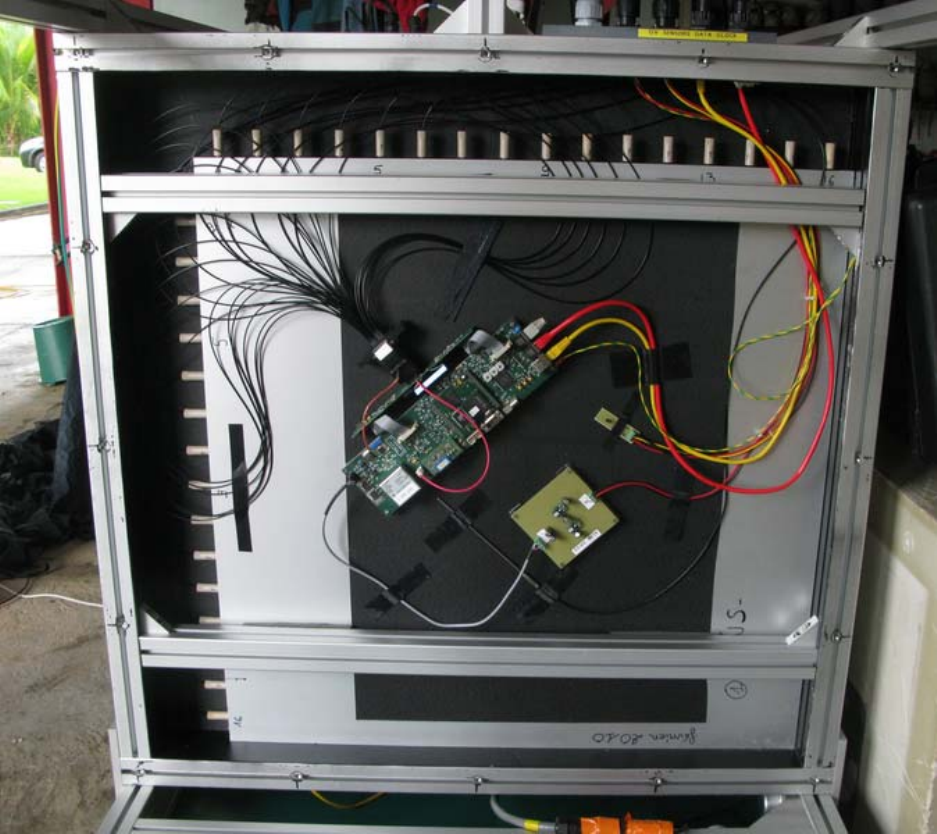}}
\caption{Left~: Picture of a muon telescope in horizontal position. The three detection planes are protected by the yellow frames. Right~: The inner part of a matrix~: active XY scintillator planes, PMT and readout electronics.}
\label{TelescopePicture}
\end{figure}

Muons ``telescopes'' are standard high energy physics trackers that count the particles arriving from a given direction during a known time slot (typically one month for the present applications). Different detection techniques are used world-wide and our telescopes are scintillator XY hodoscopes, readout by multi-anodes photomultiplier tubes (MaPMT) and self-triggered electronics, with three active planes. A picture of one telescope deployed on the slopes of the Soufri\`ere de Guadeloupe is displayed on Fig. \ref{TelescopePicture}.

The coincidence and majority logics within those three planes enable to pratically eliminate the random coincidences and to reduce the contamination from particles bundles when combined with track selection filters. The basic timing resolution step used in these telescopes is set by the master clock frequency of 20MHz which is then multiplied locally by a factor 5 via a PLL. Therefore a 10ns timing resolution was used in the first version of the telescopes to perform event timestamping and coincidence logic between the three active planes. Although sufficient to perform standard structural imaging, this timing resolution was shown to be quite limited for more sensitive studies and should have been upgraded to reach the sub-nanosecond domain. In particular we showed in the early data taking on Mount Etna and on La Soufri\`{e}re of Guadeloupe (Lesparre et al. 2012b, Jourde et al. 2013) that muons propagating downwards, from below the horizontal plane, and passing through the telescope may mimic muons that would have crossed the volcano. Indeed in both cases the telescopes were located on the steep flanks of the volcanoes and their rear-side was facing large and deep valleys allowing even low energy particles to reach and cross the telescope.
The only way to remove this noise it to use high-resolution {\it tof} criterion and therefore improve the timing resolution by at least one order of magnitude. The paper is organized as follows. In the first Section is described the DAQ and timing system of the telescopes. In the second Section the general features of the TDC implementation in the FPGA are detailed. In the third Section we present the calibration procedure and the corrections obtained on the measured flux when the backward contribution is subtracted.

\section{Description of the distributed DAQ and timing systems}

\subsection{General features}
Each detection plane is read out by its own Ethernet-capable autonomous electronics and sends its data after pre-processing to a central, low-power consumption, rugged computer for the global event building and track selection. The computer (Ref.MMI-4087AD-R2A18320 from KEP, 1.8GHz processor running under Linux) is hosted in a compact and isolated ``electronics box''. This bow also includes a network switch, low voltage regulators, a ten-relay module with a built-in web server (WebRelay-10 from Xytronix Research \& Design, Inc.) and the Master Clock board (see below).\\
The readout electronics system is directly derived from the one of the OPERA neutrino experiment(Acquafredda et al. 2009; Aganofovna et al. 2013). This experiment aimed at detecting the $\nu_\mu \rightarrow \nu_\tau$ neutrino oscillations between the CERN (Geneva, Switzerland) and the Gran Sasso underground laboratory (L'Aquila, Italy). The Target Tracker of the OPERA detector, a scintillator hodoscope, was meant to build the trigger signal and to identify the volume inside the detector where the neutrinos interacted.\\
The DAQ system, common to the OPERA and to the muons telescopes, has been designed on the concept of the so-called ``smart sensors" in which each detecting element is readout by an independant micro-processor board and sends its data over a standard Ethernet network (Marteau 2010). The DAQ system fulfills the requirements of triggerless operation, event timestamping (with 10ns steps) locked on external clock for off-line correlation, continuous running capability, high availability and low deadtime, modular and flexible hardware/software architecture (trigger schemes, on-line and off-line filters etc).

\subsection{Electronic readout system}

The muons energy loss in a scintillator strip (Pla-Dalmau et al. 2001) is converted into scintillation photons trapped in a wavelength shifting optical fibre and directed on one MaPMT pixel, where it is converted into a charge signal with a typical gain of 160fC per photo-electron. The readout electronics is divided into a front-end stage (two dedicated 32 channels ASICS developed in LAL, Orsay, France), a mother-board and a mezzanine board (constituting the hereafter so-called controller board).

\begin{figure}[!ht]
\centerline{\includegraphics[height=7cm,width=0.6\linewidth]{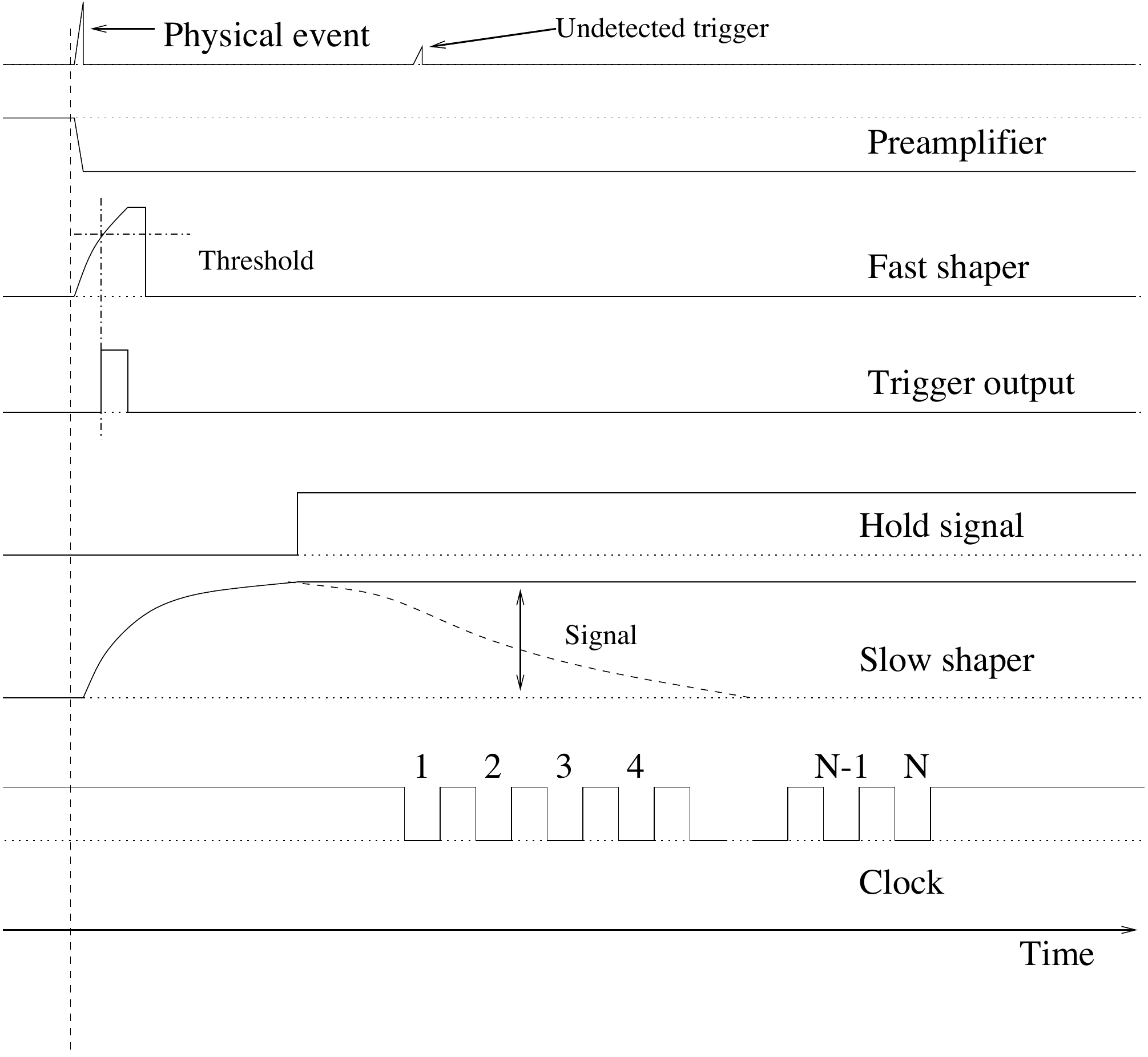}}
\caption{Timing diagrams of the front-end electronics ASICS readout sequence.}
\label{fig:timing}
\end{figure}

\paragraph{Front-end electronics} The ASICS developed is based on a standard slow/fast shaper architecture. Each channel starts with a gain-adjustable preamplifier which feeds a fast shaper, followed by a discriminator, and a slow shaper, used for precise charge readout. If an event occurs (signal above threshold on one channel's fast shaper) a track\&hold logic blocks the charges in the 32 channels slow shapers after an adjustable delay. The 32 values of the charges are then multiplexed and sent to the controller board for further processing (digitization, zero suppression, formatting and compression).
The timing processes are described in Fig.~\ref{fig:timing}.

\begin{figure}[!ht]
\centerline{\includegraphics[height=6cm,width=0.6\linewidth]{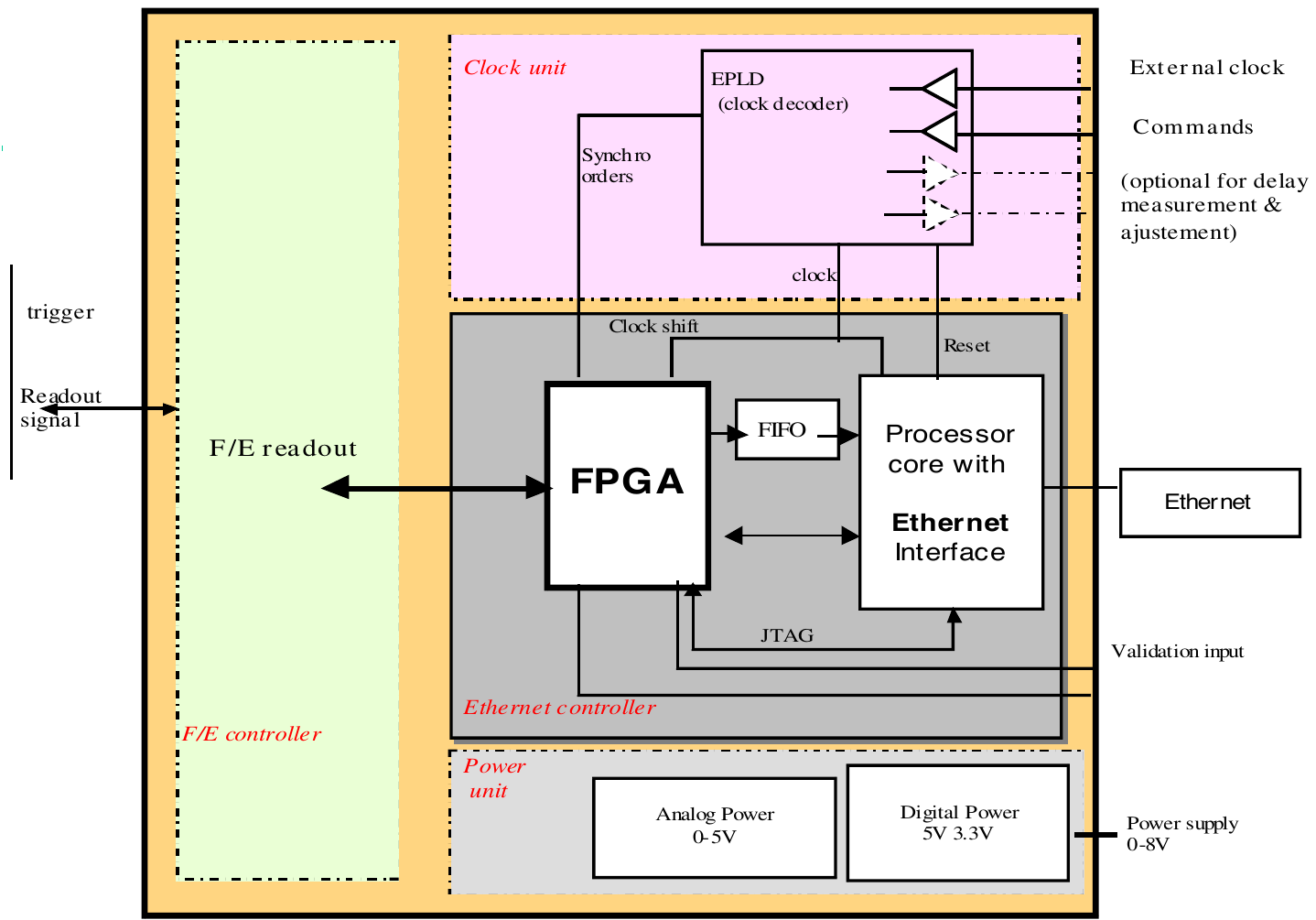}\hfill
\includegraphics[height=5.5cm,width=0.3\linewidth]{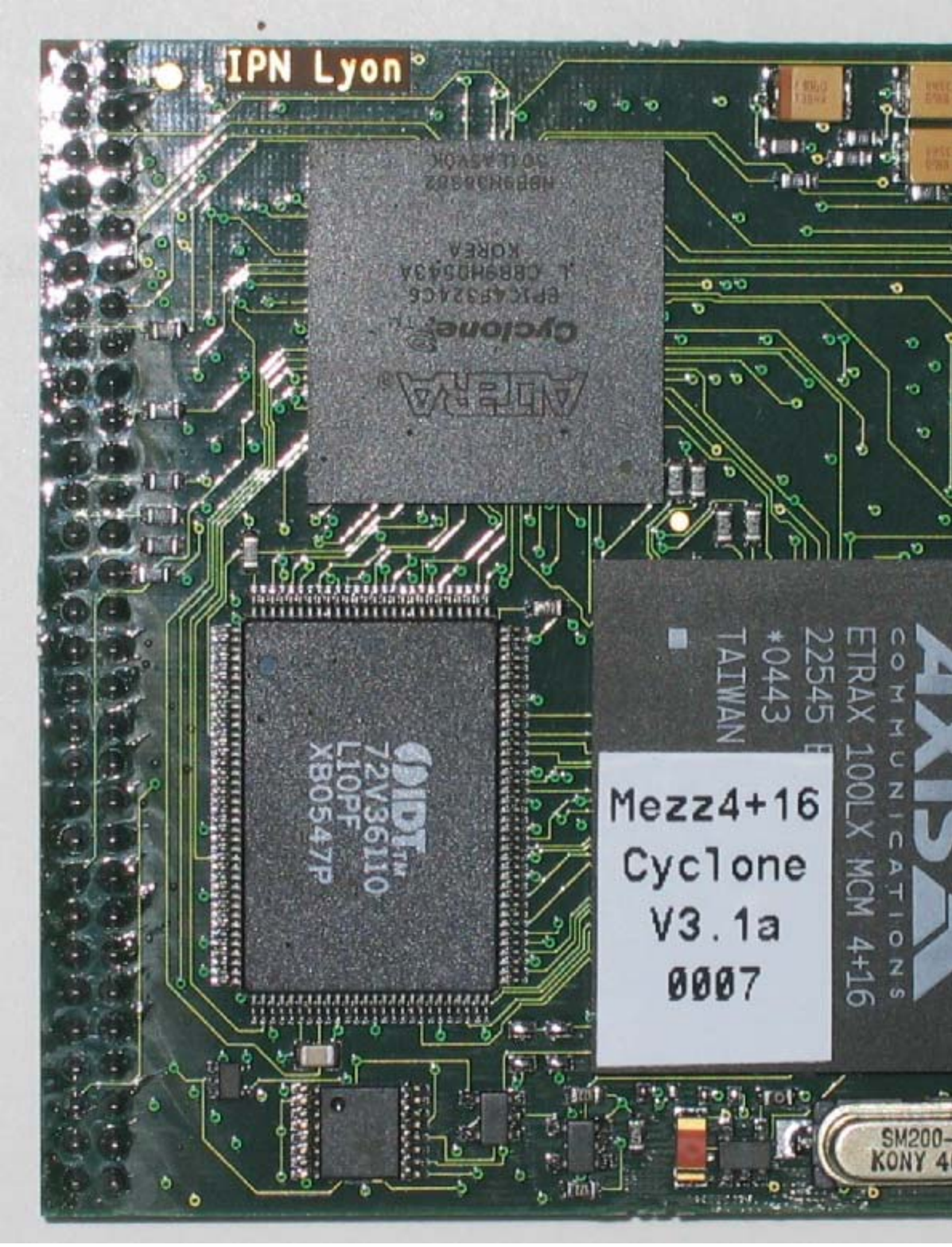}}
\caption{Left~: generic blocks diagram of a controller board (motherboard + mezzanine). Right~: picture of a mezzanine board, typically 6cm $\times$ 6cm.}
\label{fig:scheme}
\end{figure}

\paragraph{Processor boards} The mezzanine board hosts three main elements:
\begin{itemize}
\item A micro-processor, the ETRAX 100LX from AXIS, which is a 32-bit RISC CPU with Linux 2.4 OS running at 100MHz and presented in a Multi Chip Module (MCM) with 4 Mbytes of flash memory, 16 Mbytes of SDRAM and Ethernet transceiver.
\item A FPGA, ALTERA EP1C4F324C6 from the low-cost cyclone series.
\item An intermediate buffer, the 131kwords FIFO 72V36110 from IDT.
\end{itemize}
The mezzanine is designed to interface and control the F/E electronics, to handle the readout sequence, sort the data in time and transfer them to the event builder computer. It also manages the slow control processes. All communications are performed through Ethernet, see Fig.~\ref{fig:scheme}.
The readout sequence starts with the occurence of an event trigger. The FPGA generates a timestamp on this trigger (see below), performs some control logic if required (e.g. fast coincidence between two different chips) and sets the track\&hold logic. Multiplexed data are sent to the ADC (the maximal processing frequency is 5MHz which leads to an overall 6,4$\mu$s dead-time for 32 channels. The ADC values are zero suppressed, formatted, stored into the FIFO and finally passed to the micro-processor. The data sharing between the FPGA and the micro-processor is performed via IRQ generation either on a PPS basis or on an adjustable threshold on the FIFO occupancy.

\paragraph{Motherboards} The motherboard hosts, on top of the mezzanine board, the following elements:
\begin{itemize}
\item A high-voltage module (ISEG BPn-10-165-12K), with 12V input voltage and output voltage in the [0-1kV] range.
\item A second FPGA (ALTERA EPM3256ATC) dedicated to the clock management.
\item A LED pulser board which generates fast pulses to test the PMT response.
\item RJ45 connectors for data and clock.
\end{itemize}
The data acquisition scheme described here is fully auto-triggered, as required by the field applications, and only fed by low voltage signals, for a total consumption of less than 8W per detection plane, at full load.

\subsection{DAQ architecture and clock distribution}

The DAQ system is built like a standard Ethernet network, each node of which is connected to a sensor (i.e. a detection plane) or the central event-building computer. This organization gives a maximal flexibility since adding a new detection plane is as simple as connecting another RJ45 cable to a switch. All network elements used in our application are commercial ones and low-cost.\\
Since the network is totally asynchronous, an additional signal should be sent to the sensors for synchronization purposes. A derived version of the OPERA clock system has been adopted here. A common clock board (hereafter referred to as the Master Clock or MC) is embedded in the ``electronics box'' and delivers a 20MHz CLK signal and a PPS signal. These signals are encoded and decoded by M-LVDS transceivers (MLVD202 from Texas Instruments), which enable to send specific commands through the clock bus. The link is also bi-directionnal to ensure the reception of acknowledgement signals and the measurement of the propagation delays in the cables. The CLK and PPS signals are chained from one sensor to the other. Up-to 8 sensors may be chained in this way following the M-LVDS specifications and 2 clock bus are available on a single MC (therefore feeding up-to 16 sensors).\\
On each sensor the ``clock unit'' (Fig.~\ref{fig:scheme}) receives and processes the clock signals. The CLK signal is used to generate locally, in a PLL, a 100MHz clock signal. This signal drives all the readout sequences and the event timestamping process.\\
The PPS signal (also called DAQ cycle) resets the value of the local counter running at 100MHz. This procedure avoids the long-term drifts of local counters due to the bad quality of the local quartz oscillators embedded in the mezzanines. At the reception of a new PPS signal, a flag is set and a word is inserted in priority inside the FIFO. In this way, data are naturally time-ordered and subdivided in periods corresponding to the PPS duration. An incremental number is attributed to each DAQ cycle (with a maximal of 255, corresponding to a few minutes for a real PPS signal). The event-building computer sends requests to all sensors to check whether data of cycle $N$ are available. If not, the request is repeated until either all sensors answer or the maximal timeout value is exceeded. When data are available, the corresponding block is sent from the ETRAX micro-processor to the event-building computer. The same procedure is repeated for cycle $N+1$. If a sensor does not answer in due time, a warning is emitted to the computer, stating that data are missing for that specific cycle. This DAQ architecture has shown very efficient behaviour during the entire OPERA data taking period, from August 2006 (commissioning run) to December 2012 (last neutrinos from CNGS).

\subsection{Event timestamping}

\begin{figure}[!ht]
\centerline{\includegraphics[height=6cm,width=0.8\linewidth]{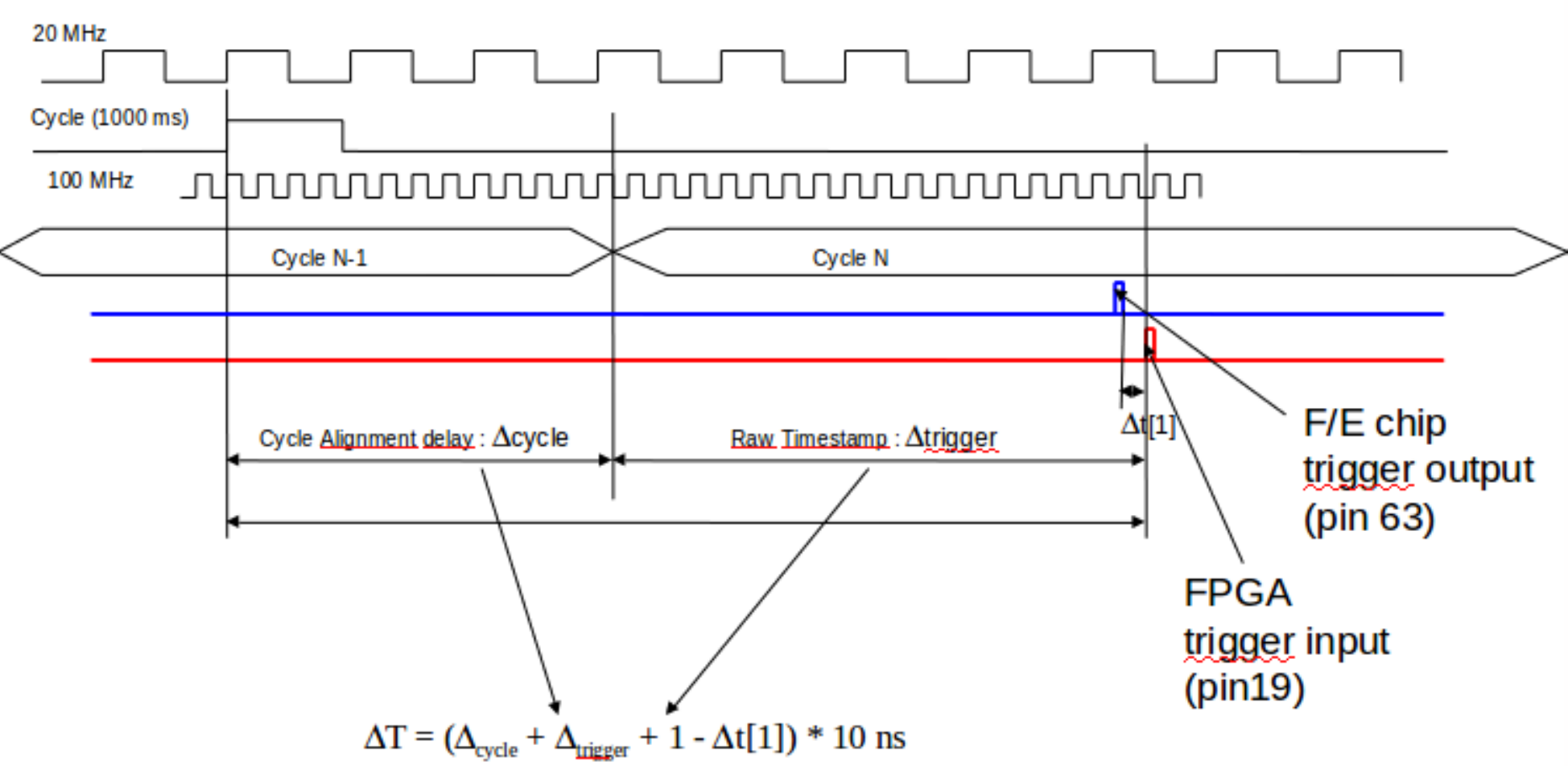}}
\caption{Timestamp generation diagram. $\Delta$cycle accounts for the propagation delays within the chained clock bus. $\Delta$trigger is the raw timestamp given in the data (ranges from 0 to 1000ms by steps of 10ns). $\Delta$t[1] is the delay for the trigger to go from front-end chip to the FPGA. The absolute time $T$ (from the starting point of the clock chain) is given by the following formula : $$(\Delta cycle + \Delta trigger + 1 -  \Delta t[1] )*10 ns <   T  < (\Delta cycle + \Delta trigger + 2 - \Delta t[1])*10 ns.$$ All delays to reach the 1st controller board in the clock chain should be added.}
\label{fig:timing2}
\end{figure}

The standard event timestamping procedure is based on the latching of the current local counter generated at 100MHz inside the FPGA, on the occurence of a trigger from the front-end electronics. As described above this launches the readout sequence (track\&hold, multiplexing, analog-to-digital conversion, Fig.~\ref{fig:timing}).\\
Two special time flags are inserted into the data flow, one for the cycle number (which is then common to all data recorded within that cycle) and one timestamp is attached to each individual event in 10ns units (the ultimate timing accuracy at this level). Since the computer links the DAQ cycle time to the UTC, the absolute timing of each event is then known by summing the cycle time value and the 10ns timestamp. The timestamp procedure and the propagation of delays are displayed in the right part of Fig.~\ref{fig:timing2}.\\
Although the 10ns accuracy is enough for standard applications, it has been already discussed that it may reveal insufficient for smarter data analysis. Of course a basic idea is to increase the CLK frequency in order to get finer timing steps. But the precision to reach, below one nanosecond, makes this option unpractical and too critical in terms of power consumption, a key feature of the unmanned sensors.\\
Since the system was not designed initially neither for precise time measurements nor for particles time-of-flight determination, no specific hardware was foreseen. A remaining interesting option is the use of TDC techniques like tapped delay lines, delay locked loop, vernier delay lines or trigger ring oscillator. Most of those techniques are implemented in dedicated ASICS which obviously was not possible here. The idea was therefore to integrate TDC modules within the existing FPGA of each mezzanine board (Lin et al. 2006; Song et al. 2006; Junnarkar et al. 2008; Junnarkar et al. 2009). This technique allows a significant improvement of the system performances, simply by reconfiguring the FPGA via hardware description language (HDL). The techniques used for this particular upgrade implied no new hardware, allowed many tests on different configurations and are easily implemented (the use of an emulated JTAG bus in the case of the sensor mezzanine between the microprocessor and the FPGA enables the reconfiguration without direct connection of a programmation tool).

\section{Upgrade of the timing system}

\begin{figure}[!ht]
\centerline{\includegraphics[height=3cm,width=0.45\linewidth]{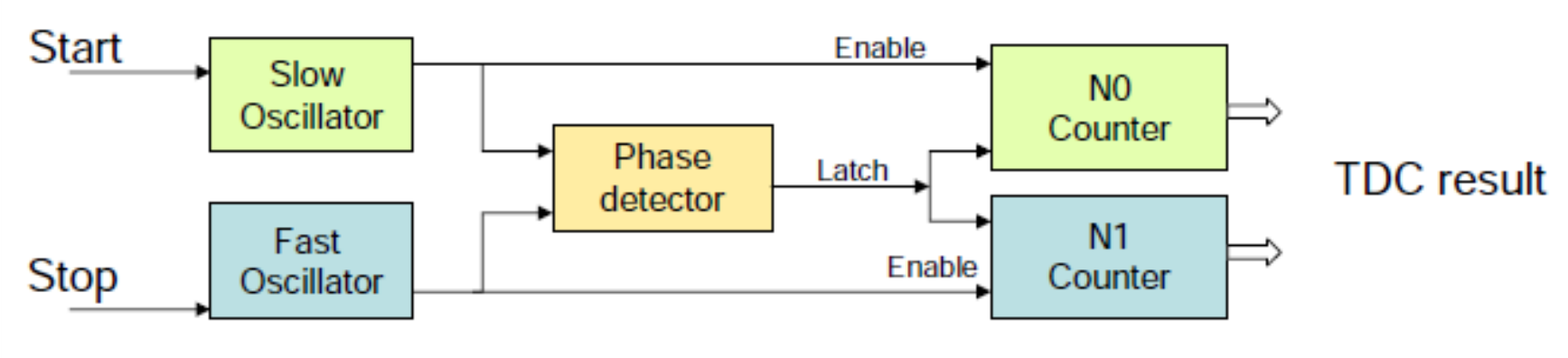}\hfill
\includegraphics[height=3cm,width=0.55\linewidth]{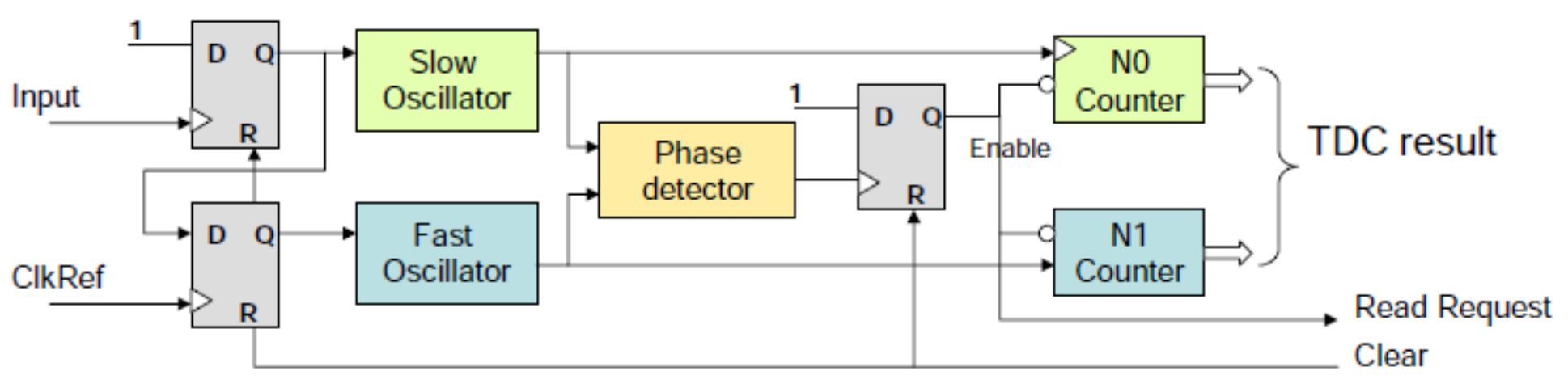}}
\caption{Left~: block diagram of a generic ring oscillator TDC. Right~: block diagram of a ring oscillator TDC with external reference clock signal used to generate the STOP signal of the TDC.}
\label{fig:tdcA1}
\end{figure}

With a standard counter the measurement of a time interval relies on the number of clock counts recorded. The intrinsic error on the measurement equals the quantization error and the resolution is limited to the clock period. The idea to overcome this limitation is to use a vernier with two slightly different and controllable frequencies $T_{slow}$ and $T_{fast}$ and to use the difference in periods as the new timing step. A phase detector is used to perform the coincidence between the two counters (Fig.~\ref{fig:tdcA1}, left). Of course, the key parameter is the generation of the two nearby clocks in a controlled and reproducible way.

\subsection{Ring oscillator TDC}
In the present applications the 100MHz clock generated on the mezzanine is used as the reference STOP signal. The physical signal under study generates the START. The schematics is displayed in Fig.~\ref{fig:tdcA1}, right. The two elements of this type of TDC system are the controllable ring oscillators and the phase detector.

\begin{figure}[!ht]
\centerline{\includegraphics[height=6cm,width=0.7\linewidth]{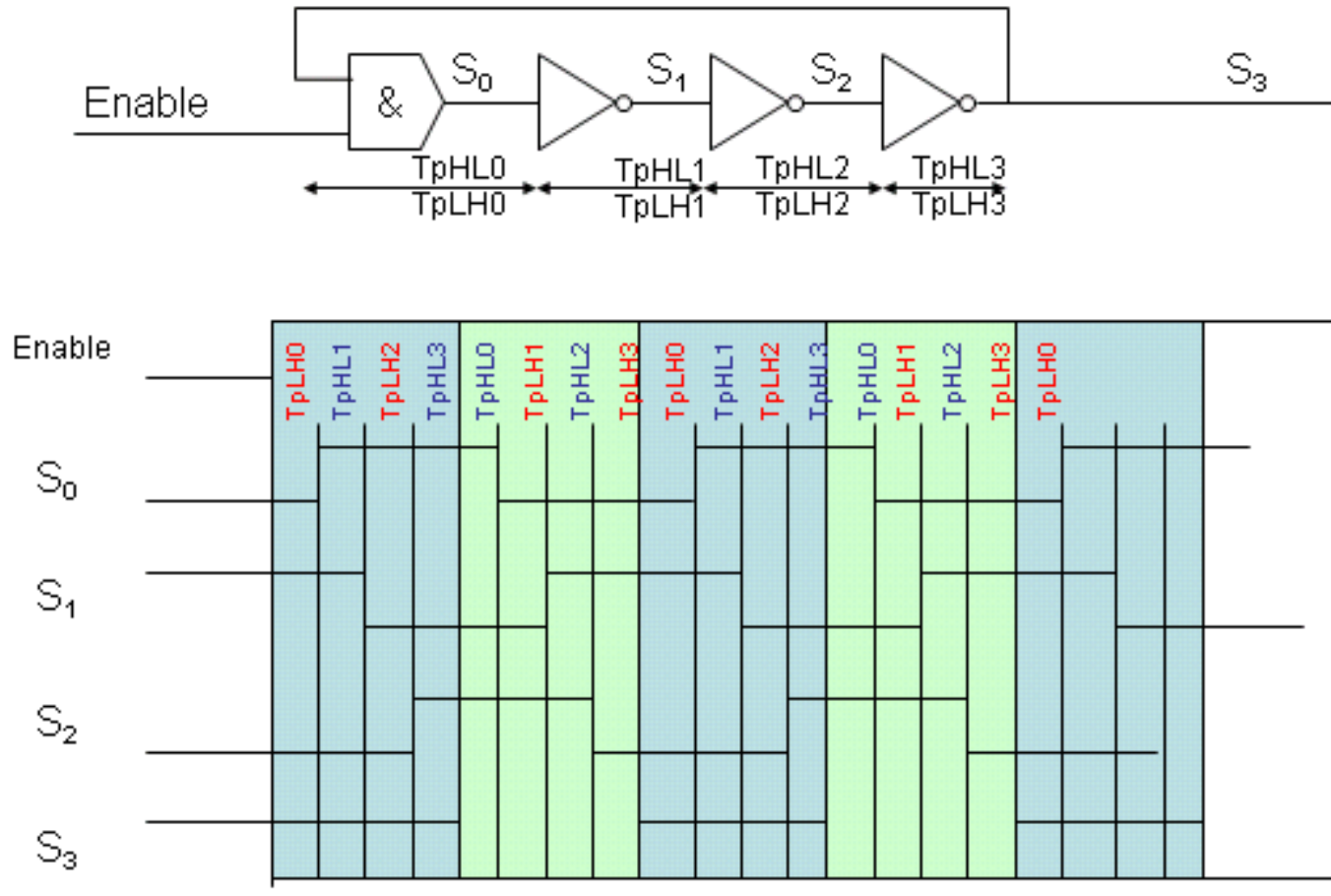}}
\caption{Simple ring oscillator and its timing diagram. $T_{pHLn}$ and $T_{pLHn}$ denote the times of the logic when it goes from state high (H) to low (L) and vice-versa.}
\label{fig:tdcA2}
\end{figure}

\paragraph{Ring oscillators} The oscillators are generated following simple schemes. An example of ring oscillator developed and validated in the present application is detailed in Fig.~\ref{fig:tdcA2}. An odd number of logical cells is required to generate the oscillation through the feedback loop. The oscillation starts as soon as the ENABLE signal flips from 0 to 1. The period of the final signal ($S_3$ in this example) depends on the delays cumulated in the logical elements themselves and in the propagation from one to the other. Given the asymmetry of the logic when it goes from high state (H) to low state (L) or in the opposite direction (from (L) to (H)), we denote the various periods respectively $T_{pHLn}$ and $T_{pLHn}$, with $n=0..3$ (see Fig.\ref{fig:tdcA2}). The final frequency is then given by the following formula~:$$f = \frac{1}{\sum_{n=0}^{n=3} T_{pHLn} + T_{pLHn}}$$
It is worth noticing that the final frequency is given by the routing within the FPGA. Therefore the way the oscillators are implemented into the FPGA may be as controlled as possible to grant a good level of reproducibility.

\begin{figure}[!hb]
\centerline{\includegraphics[height=5cm,width=0.7\linewidth]{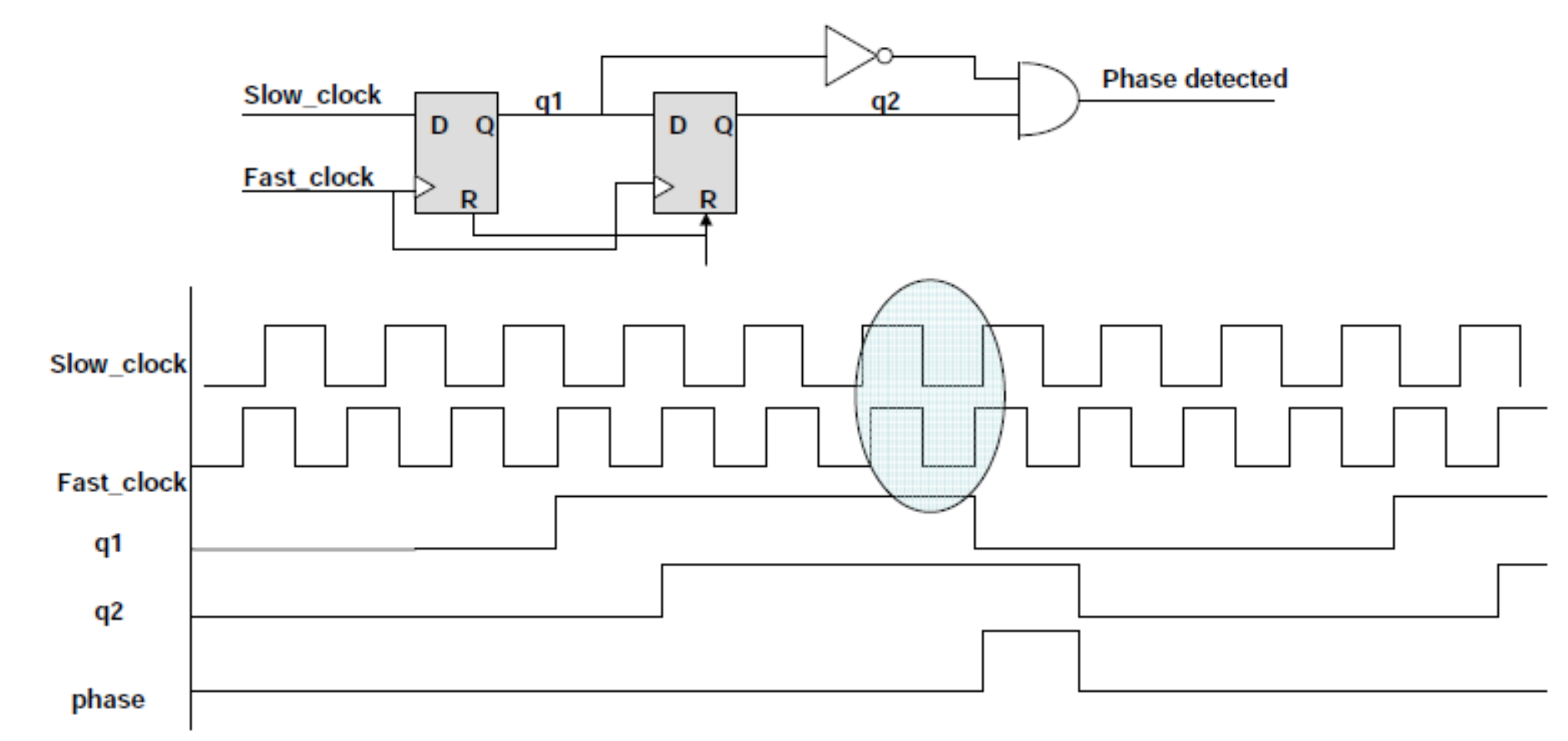}}
\caption{The phase detector and its timing diagram.}
\label{fig:tdcA3}
\end{figure}

\paragraph{Phase detector} The phase detector (Lin et al. 2006) developed and validated for this application is displayed in Fig.~\ref{fig:tdcA3}.
Also shown is the timing diagram of this phase detector. The phase detector is actually the most limiting part of the system to reach very small time resolution because of the setup time of the flip-flops. They may be left in metastable positions and require some time to recover.

\subsection{Implementation in FPGA}
Since the idea is to generate two oscillators with close frequencies due to small difference in the FPGA routing, it is worth using special procedures to freeze the design in a controlled way. Indeed a direct synthesis may give random results and differing from one synthesis to the other. The target being the tens of pico-seconds time resolution, a careful design is mandatory.\\
Using standard design tools available from ALTERA, the following design sequence has been used in this project~:
\begin{itemize}
\item creation of a separate project for the ring oscillators
\item creation of regions for each oscillator
\item open loop synthesis with SDC constraints
\item Timequest analysis
\item Post compilation Editing (ECO) with chip planner
\item regions export
\end{itemize}
In particular the regions should be located to their final position in the TOP project, and the in/out pins should be inserted manually. During the final regions import, the netlist-type should be set to Post-fix in order to protect the region. Following this design sequence allows to generate, in a reproducible way, ring oscillators with close frequencies to achieve TDC measurements.

\subsection{Timing sequence}
\begin{figure}[!hbp]
\centerline{\includegraphics[width=0.7\linewidth,height=4cm]{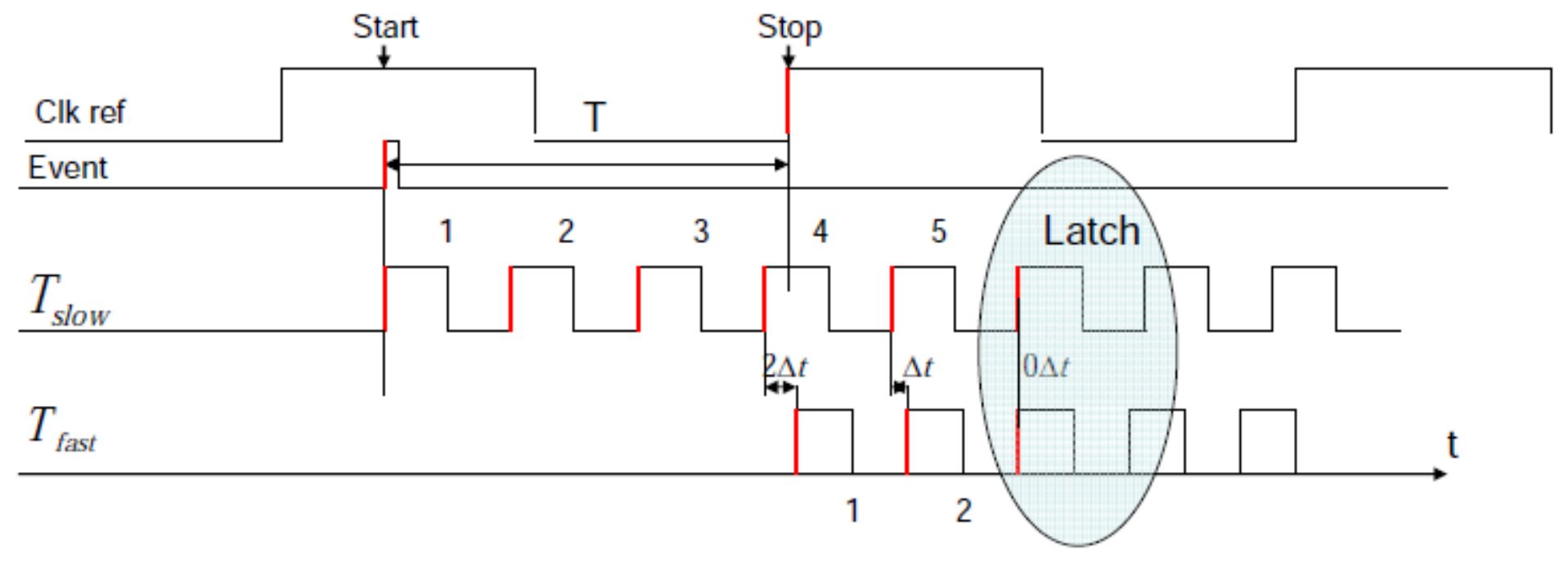}}
\caption{TDC vernier principle.}
\label{fig:vernier}
\end{figure}

The ring oscillators, called ``fast'' and ``slow'' oscillators have period $T_{fast} = T_1$ and $T_{slow} = T_0$ respectively ($T_0 \geq T_1$). The slow oscillator is launched on the rising edge of the signal trigger while the fast one starts on the rising edge of the local 100MHz clock. The timing sequence is summarized in Fig.~\ref{fig:vernier}.

The determination of the coincidence, using the phase detector, between the two signals gives access to two values, $N_0$ and $N_1$, representing the number of counts for the slow and the fast clock respectively. The timestamp of the signal, wrt the local counter, is therefore given by: $$ T = (N_0 T_0 - N_1 T_1) =  (N_0- N_1).T_0 + N_1.\Delta t ,$$
In the previous equation $$\Delta t = T_{0} - T_{1}$$ represents the time resolution of the system. The deadtime of the system depends on the ratio of the two quantities~: $T_0 / \Delta t$.\\
Obviously the 2 parameters to be calibrated are $\Delta t$ and $T_{0}$. Typical values obtained for these parameters in the ALTERA EP1C4 FPGA of the mezzanines are $\Delta t \sim 0.2$ns and $T_{0} \sim 2.2$ns. This leads to a factor $\sim$50 improvement in the time resolution wrt the 10ns steps of the original design. In the next Section we detail the calibration procedures and the results obtained in the physics analysis.

\section{Calibration and results}

In the previous sections we described the global DAQ and timing system developed for the DIAPHANE project of cosmic muons tomography. In particular we saw how the design of ring oscillators TDC enables to increase significantly the timing resolution of the system. It has been also pointed out that the design may be performed in a careful way and that the TDC parameters should be calibrated since they strongly depend on the routing procedures.

\subsection{Calibration procedures}
The $T_0$ value is calibrated with the help of the local reference 100MHz clock. The method consists in counting the number $N_{cal;0}$ of $T_0$ periods corresponding to $N_{ref}$ 10ns periods (labelled $T_{ref}$). Typical values are $N_{cal;0} \sim 256$ (8 bits counter). Therefore~: $$ T_0=N_{ref}\times \frac{T_{ref}}{N_{cal;0}} \quad \mathrm{ns.} $$
\begin{figure}[!htbp]
\centerline{\includegraphics[width=0.6\linewidth,height=4cm]{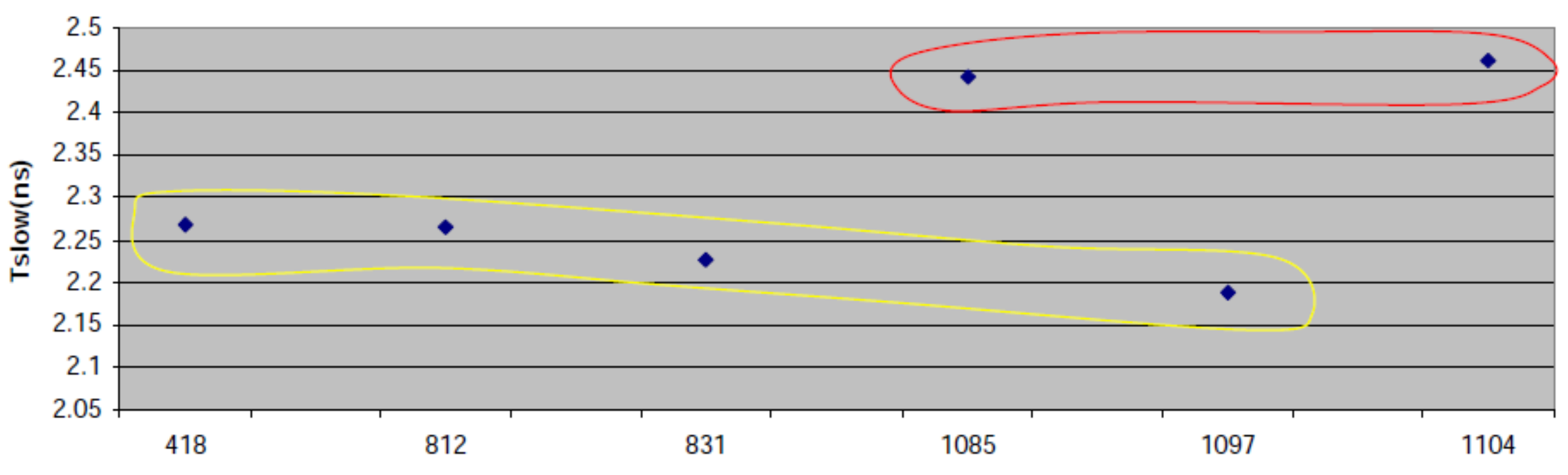}}
\caption{$T_0$ measurements on 6 different FPGA. On the X-axis we reported the serial number of the FPGA. }
\label{fig:tdc}
\end{figure}

The error on the measurement is of the order of $\frac{T_0}{N_{cal;0}}$. Results obtained on a sample of 6 different mezzanines are displayed in Fig.~\ref{fig:tdc}. We identified the effect of the different production batch on the $T_0$ value's dispersion. This result clearly shows that each sensor should be individually calibrated. \\

\begin{figure}[!h]
\centerline{\includegraphics[width=0.4\linewidth,height=3.5cm]{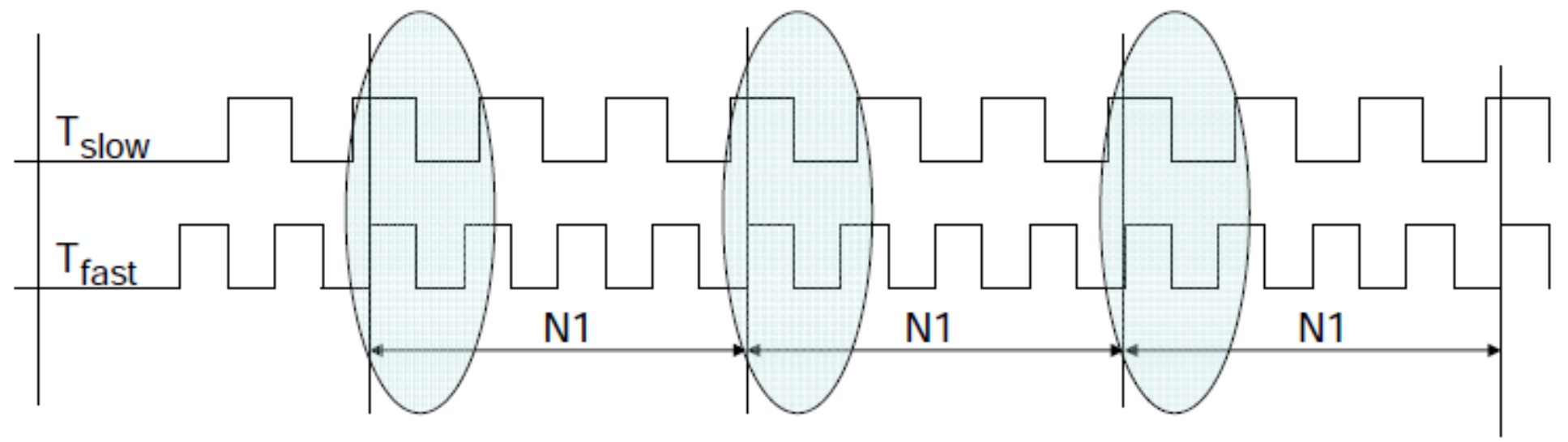}\hfill
\includegraphics[width=0.6\linewidth,height=3.5cm]{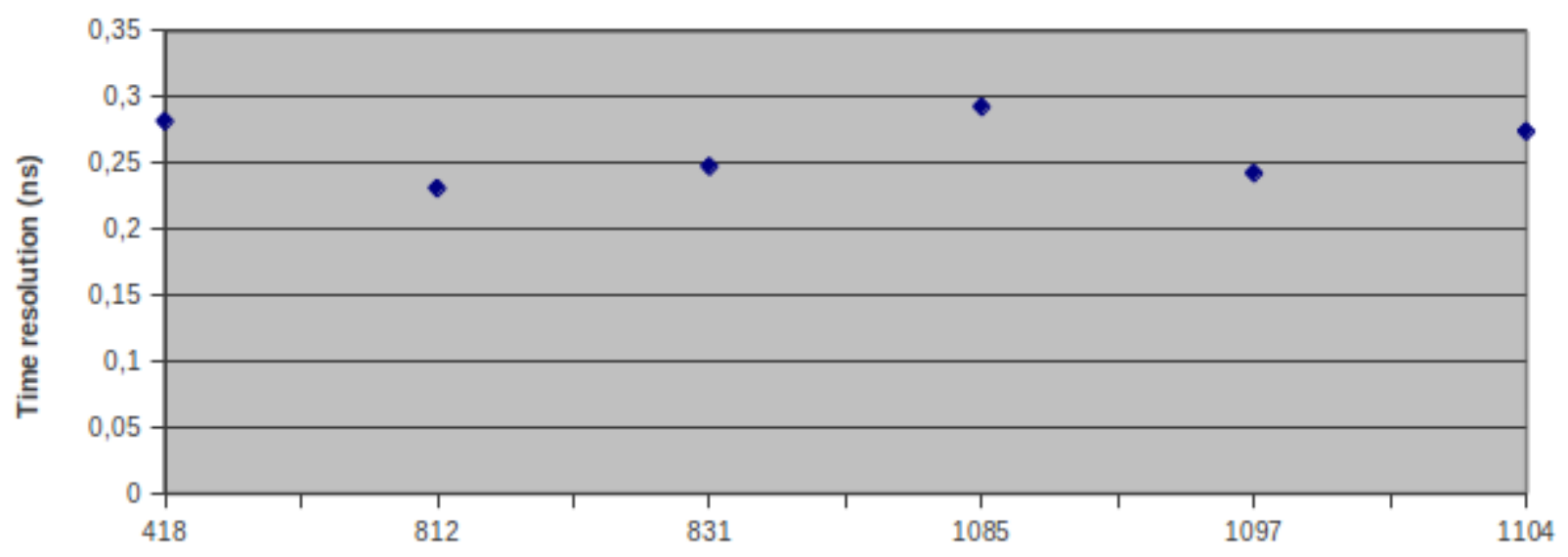}}
\caption{Left~: $\Delta t$ calibration principle. Right~: $\Delta t$ measurements on 6 different FPGA. The FPGA are the same as those used for the calibration in Fig.~\ref{fig:tdc}, identified by their serial number.}
\label{fig:tdc2}
\end{figure}

To determine the value of $\Delta t$ an internal trigger is started with the slow oscillator and the fast oscillator starts on the next rising edge of the 100MHz clock. Then we count the number $N_{cal;1}$ of $T_0$ periods between two consecutive phase detections.
The measurements are repeated and the average value of $N_{cal;1}$ is computed. Through this procedure we get the desired value of the timing resolution from the following formula~: $$\Delta t = \frac{T_0}{\bar{N_{cal;1}}} \mathrm{ns}.$$ Results are available from Fig.~\ref{fig:tdc2}.

\subsection{Linearity tests}
\begin{figure}[!htbp]
\centerline{\includegraphics[width=0.5\linewidth,height=4cm]{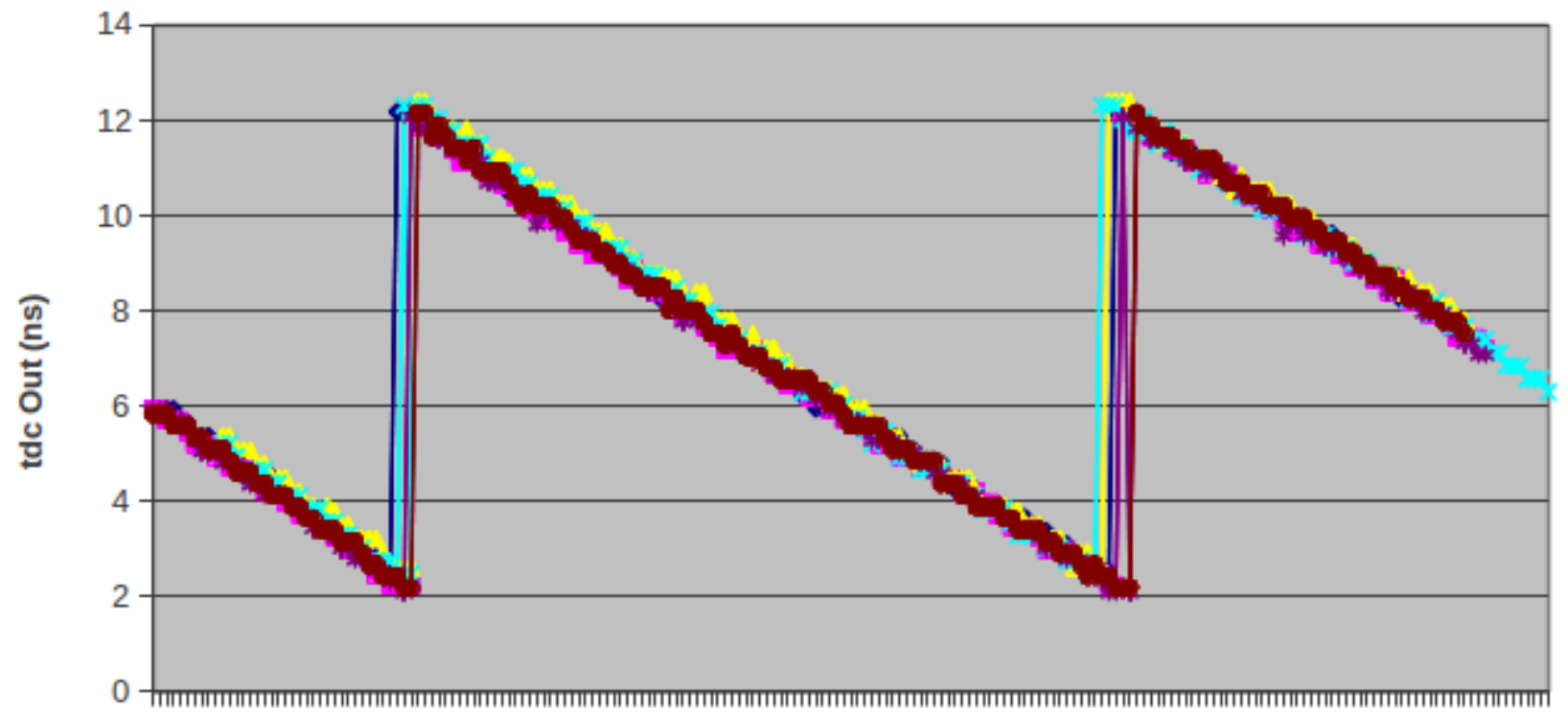}\hfill
\includegraphics[width=0.5\linewidth,height=4cm]{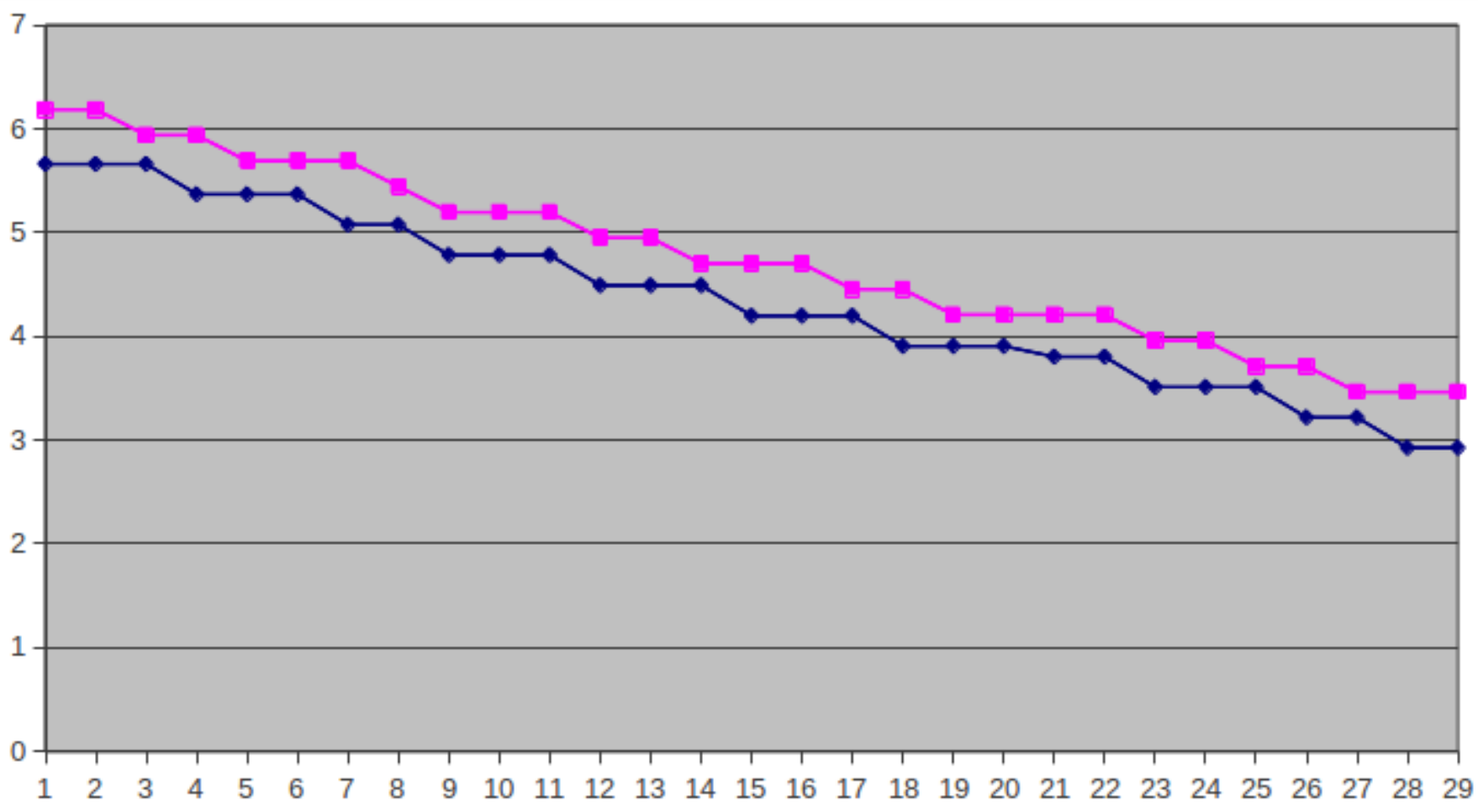}}
\caption{Left~: calibration curves for the same set of 6 sensors. The Y-axis represents the TDC output, while the X-axis corresponds to the external delay steps. Right~: zoom a measurement sequence for 2 different sensors. The offset between the two is clearly visible. The 'plateau' observed and lasting 2-3 external delay steps are compatible with the measured 240ps.}
\label{fig:linearity}
\end{figure}

\begin{figure}[!htbp]
\centerline{\includegraphics[width=0.4\linewidth,height=4cm]{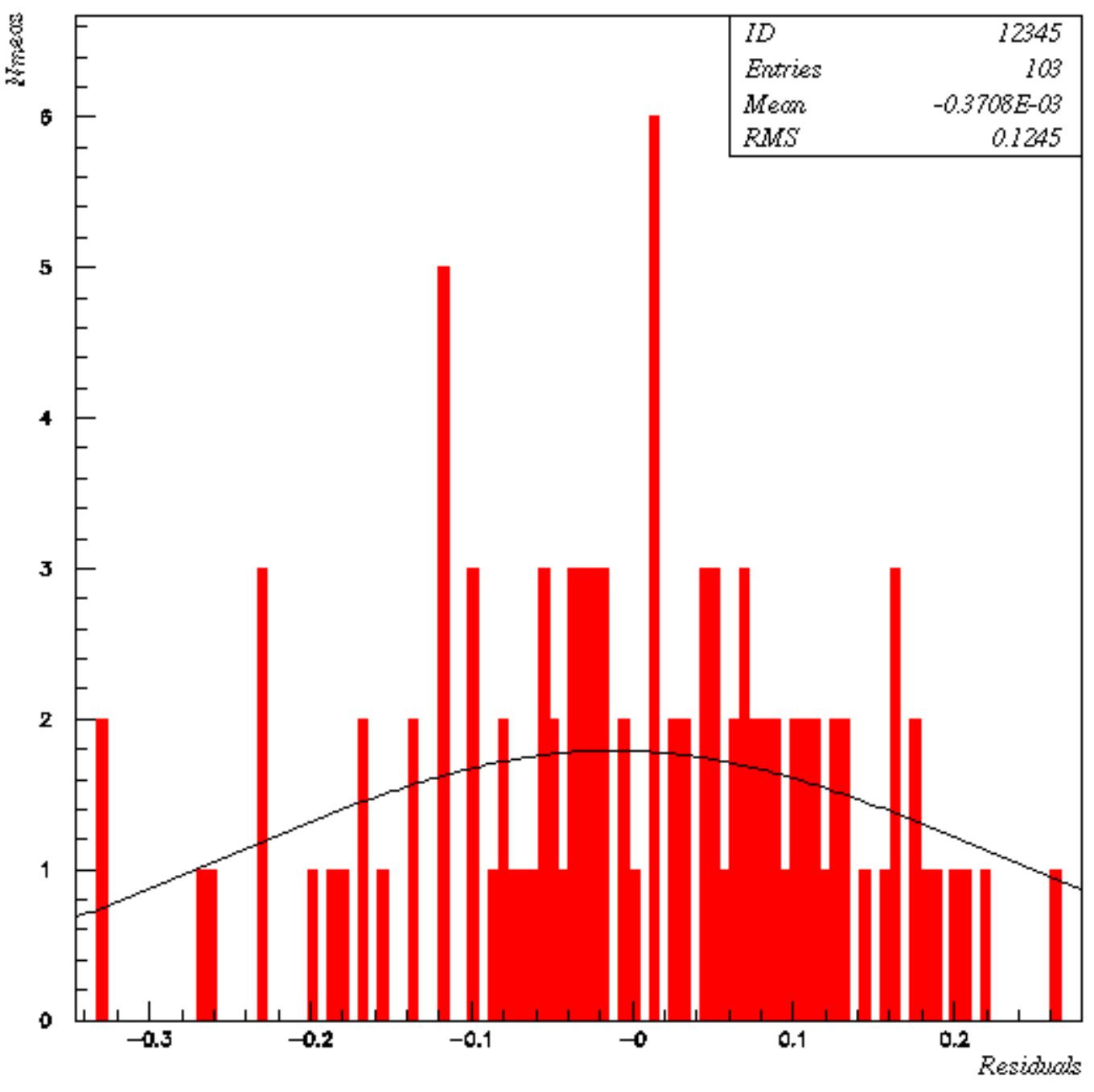}}
\caption{Linear fit of the calibration curves: distribution of the residuals.}
\label{fig:linearity2}
\end{figure}

We tested the linearity of response of single sensors by using an external delay unit (HP 8131A, adjustable steps of 10 to 100ps). The tests procedure is the following. A signal trigger is generated in the FPGA, synchronous with the 100MHz clock, using the LED PULSE MODE of the system. This channel is usually devoted to the generation of a fast pulse, synchronous to the local clock, to be sent to a blue LED illuminating the PMT. This procedure enables to measure the PMT photo-electrons response and therefore its gain.\\
To test the linearity of the TDC response we take out the generated trigger and feed an external delay unit with a resolution of 100ps. The delayed signal is then sent back to the sensor as an external trigger (a dual LEMO connector is avalaible on the motherboard). The TDC starts on the rising edge of this external trigger and stops on the next rising edge of the 100MHz. This gives the position of the received trigger inside the 10ns period. A READY status flag and a CLEAR signal manage the current measurement sequence and enable the next measurement. Between two consecutive measurements the external delay is increased by steps of 100ps. Measurements are repeated over more some 10ns local clock periods. Typical calibration curves for 6 different mezzanines are displayed on Fig.~\ref{fig:linearity}.
The sign of the slope in a 10ns period depends on the way the START and STOP are generated. The linearity of the TDC is clearly seen from the curves, all compatible with an average time resolution of 240ps. Also visible, the differences in offsets from one sensor to the other, which have to be corrected in a multi-TDC system implementation. The residuals of a linear fit to one calibration curve are displayed in Fig.~\ref{fig:linearity2}.

\subsection{Comparative tests between sensors}
\begin{figure}[!hb]
\centerline{
\includegraphics[width=0.8\linewidth,height=4cm]{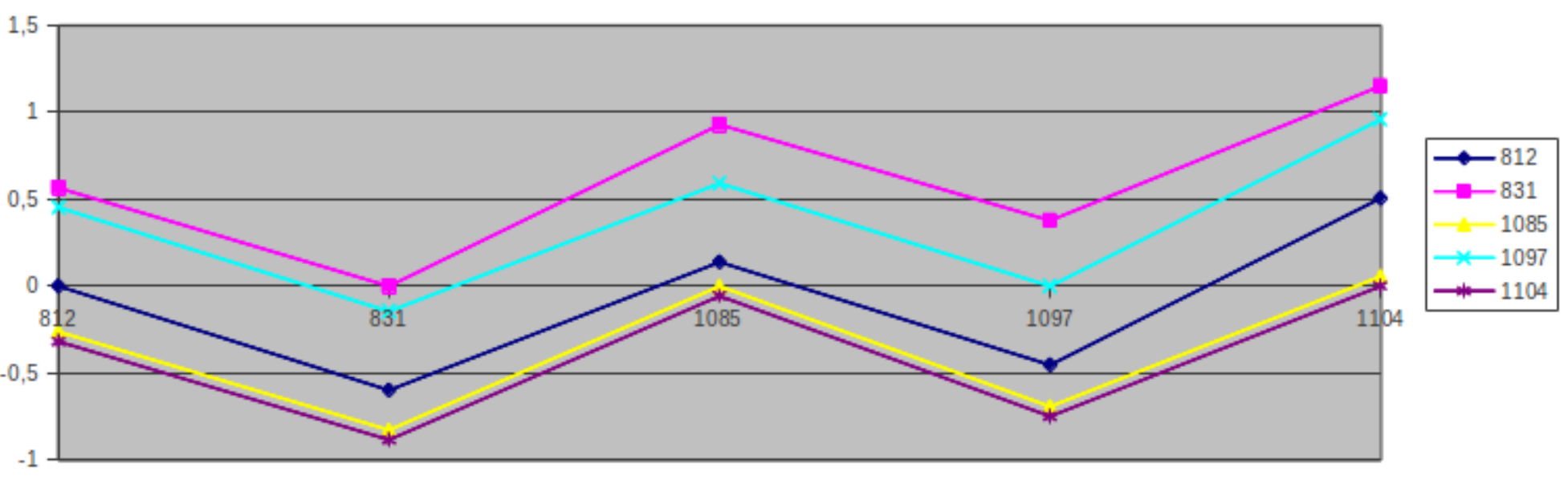}}
\caption{Systematic errors (in ns) in a series of measurements between the raw TDC output of two different sensors.}
\label{fig:comp}
\end{figure}

We use simultaneously two different sensors. The first one (Master Sensor) works in LED PULSE MODE like in the previous test. This signal is delayed externally and then resent, as EXT TRIGGER, to the two sensors. All cables paths are identical. The 20MHz clock signals from the Master Clock are transmitted by Ethernet cables with same length. This test outlines the dispersion effects between two sensors in a single measurement sequence. The systematic errors on a series of measurements for each combination of two sensors are reported in Fig.~\ref{fig:comp}.

The effect of the FPGA's batch production is particularly sensible for serial number 1085 and 1104, belonging to same batch, and which curve shows quasi identical responses wrt to each other and to the four remaining sensors. This result points out the requirement of a complete calibration procedure, as described in the preceeding subsections, to fully exploit the fine resolution of the TDC.

\subsection{Results obtained on field}
Detailed results are given in (Jourde et al. 2013). In the following Fig.~\ref{fig:tof} the time-of-flight distribution of recorded muons as measured on a three-planes telescope. The most accurate TOF value is computed as the difference in the TDC outputs of the 2 extreme sensors because of the large lever arm. The information of partial TOF, derived from the information of the middle plane is included in the analysis as well since it provides an independant set of data to be mixed with the main one.\\
Propagation delays (in the signal cables and in the optical fibres from the scintillator to the PMT's) are corrected for. Dedicated calibration tests have been carried out for those corrections, already at the time of the OPERA experiment (Marteau 2009).\\
The telescope was set on the top of a small hill without any matter on any side. The slow slope of the hill allowed particles also to propagate upwards from below the horizon. Left part of the figure describes all variables and terms used in the analysis. The upwards propagating muons correspond to negative values of the zenith angle if they propagate backwards, and positive values if they propagate forwards.
\begin{figure}[!hbtp]
\centerline{\includegraphics[width=0.5\linewidth,height=5cm]{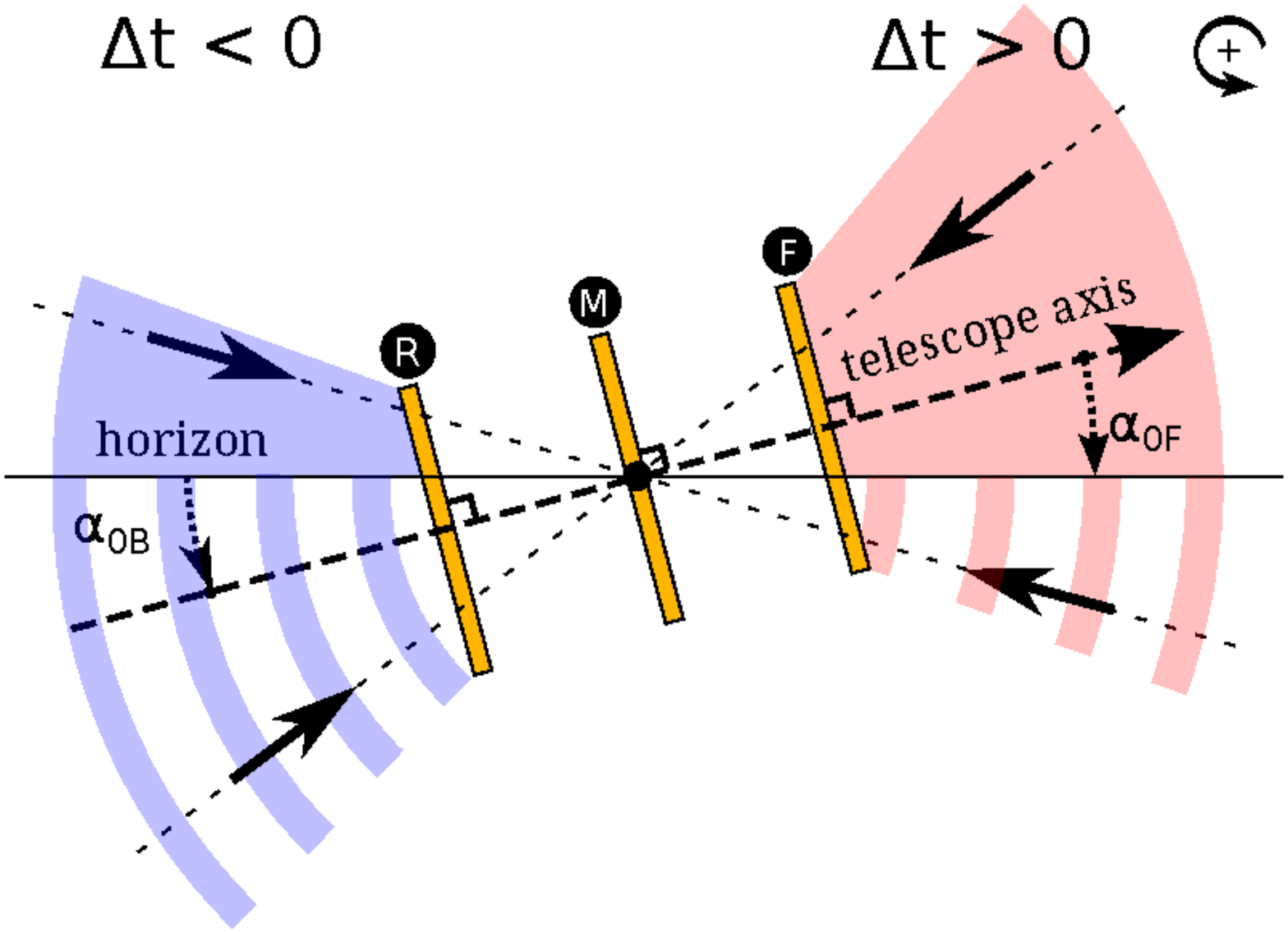}\hfill
\includegraphics[width=0.5\linewidth,height=7cm]{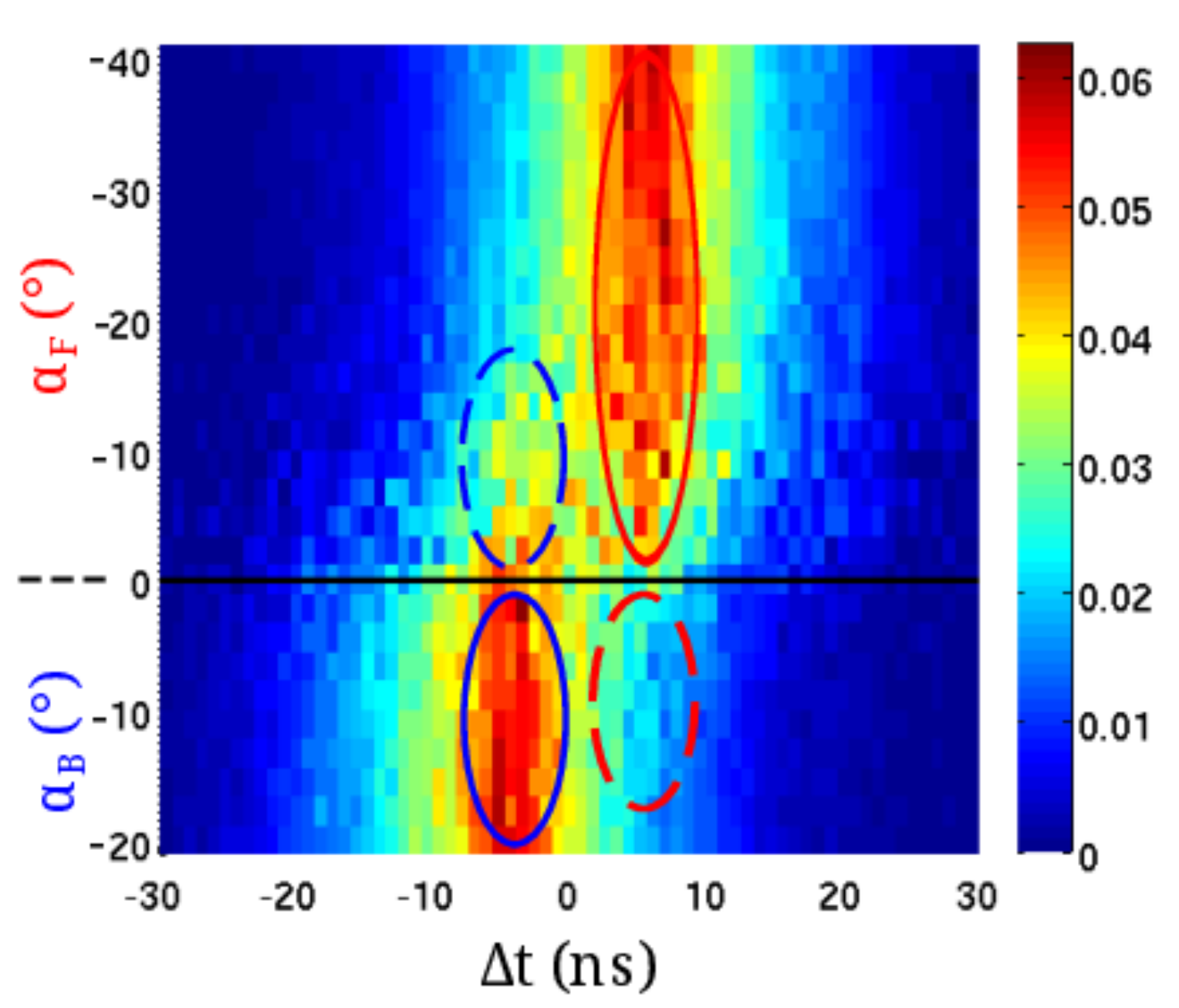}}
\caption{Left~: orientation convention for the muons propagation. The telescope is seen from the side. Right~: time-of-flight distribution wrt to the zenith angle. The horizon is represented by the dashed line. The blue and red solid ellipses respectively show the backward ($\alpha_B < 0$ and $\Delta t < 0$) and forward ($\alpha_F < 0$ and $\Delta t > 0$) events corresponding to the downward fluxes. The dashed ellipses show events corresponding to upward-going muons from forward (red ellipse, $\alpha_B < 0$ and $\Delta t > 0$) and backward (blue ellipse, $\alpha_F < 0$ and $\Delta t < 0$).}
\label{fig:tof}
\end{figure}
Of course, given the properties of the cosmic muons flux which is maximal at the zenith, the level of backwards propagating muons is expected to be rather low. The 2D-histogram shows the TOF distribution as a function of the zenith angle. The solid ellipses correspond to downwards propagating muons (the main component of the flux). The dashed ellipses clearly point out the existence of an upwards propagating muons sample that are identified thanks to the fine-grained TDC implemented in the sensors and calibrated following all the procedures described in the text.
This result allows to correct the images obtained for a large geophysical structure, like a volcano, where the lowest part of the image corresponds to the largest amount of matter to be crossed by the muons, therefore to the lowest fluxes and also to the closest-to-horizon propagation directions. If an upwards-going muon is misidentified, it may be falsely accounted for a crossing-through muon. This results in a wrong evaluation of the global volcano density which then seems less dense than in reality. This background source is extremely important and has to be reduced as much as possible for the small values of the fluxes or when one addresses physical situation where small differential fluxes may be measured. All details may be found in (Jourde et al. 2013).

\section{Conclusions and perspectives}

\subsection{Adjustable TDC}
An upgrade of the present design has been originally developed after this application. The idea is to change the oscillators configuration online to adjust their frequency and then combine different slow/fast oscillators to get different resolution. In this approach, optimal resolution may be designed for any application. Technically one varies the position of the cells and the routing. When the design is completed a calibration of all possible oscillators combinations is mandatory. One gets a catalog of TDC's with different resolutions.
\begin{figure}[!hbtp]
\centerline{\includegraphics[width=0.45\linewidth,height=4cm]{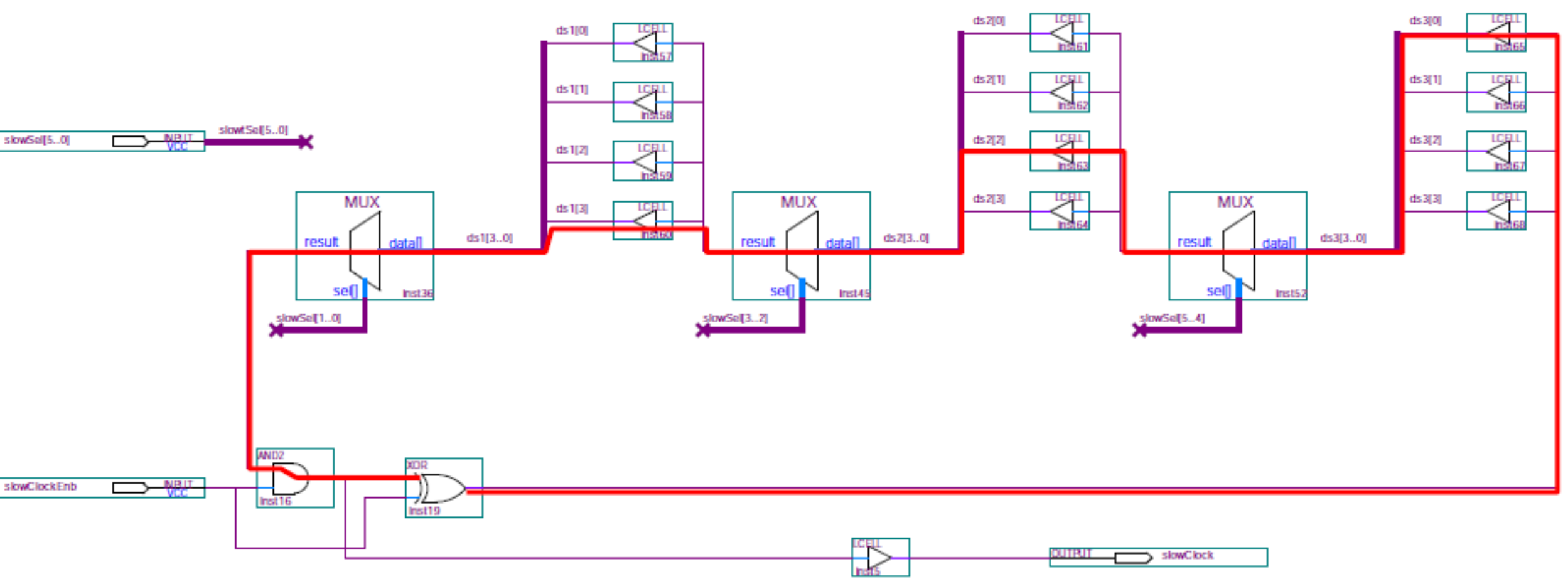}\hfill
\includegraphics[width=0.45\linewidth,height=4cm]{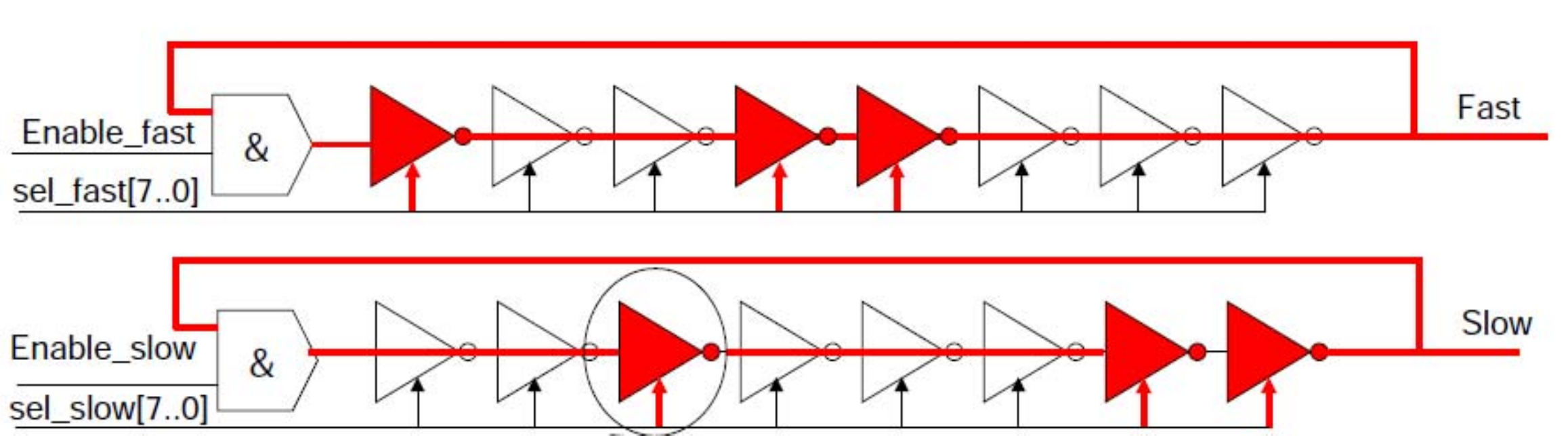}}
\caption{Left~: adjustable oscillators with MUX architecture. Different combinations of the LCELL result in different oscillation frequency, therefore in different resolutions for the TDC. Right~: adjustable oscillators with moving inverters architecture. The LCELL are selected or not in the chain.}
\label{fig:adj1}
\end{figure}

As an example we detail the schemes of two possible designs for adjustable oscillators. The first one (Fig.~\ref{fig:adj1}, left) uses multiplexors that select between different LCELL. In this example $4^3 = 64$ different oscillators may be obtained. The second example (Fig.~\ref{fig:adj1}, right) implements an architecture with moving inverters. In this architecture one benefits from the fact that the delay between passing and inverter functions are different and so are the delays between inverters. The resolution of the different combinations is computable a priori in this design by passing one XOR from passing to inverted and vice-versa.
\begin{figure}[!htbp]
\centerline{\includegraphics[width=0.45\linewidth,height=4cm]{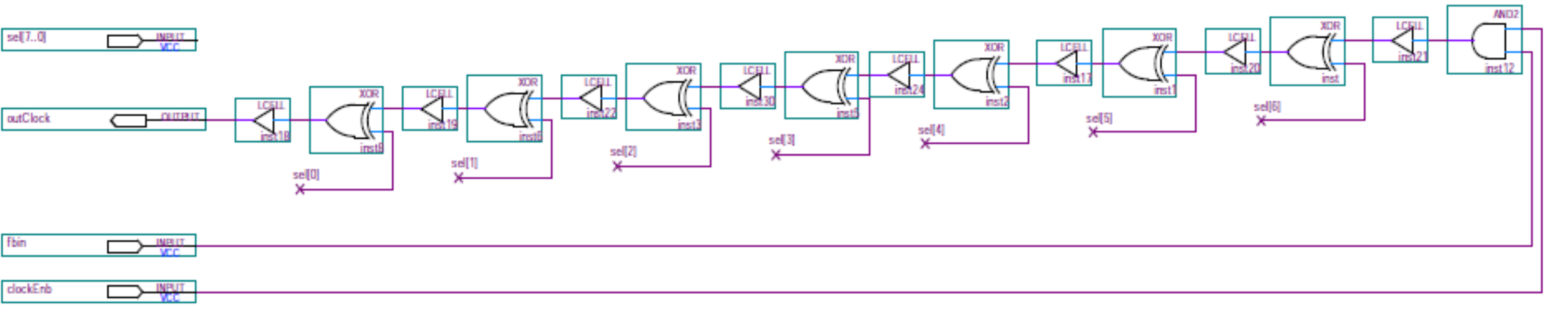}\hfill
\includegraphics[width=0.45\linewidth,height=5cm]{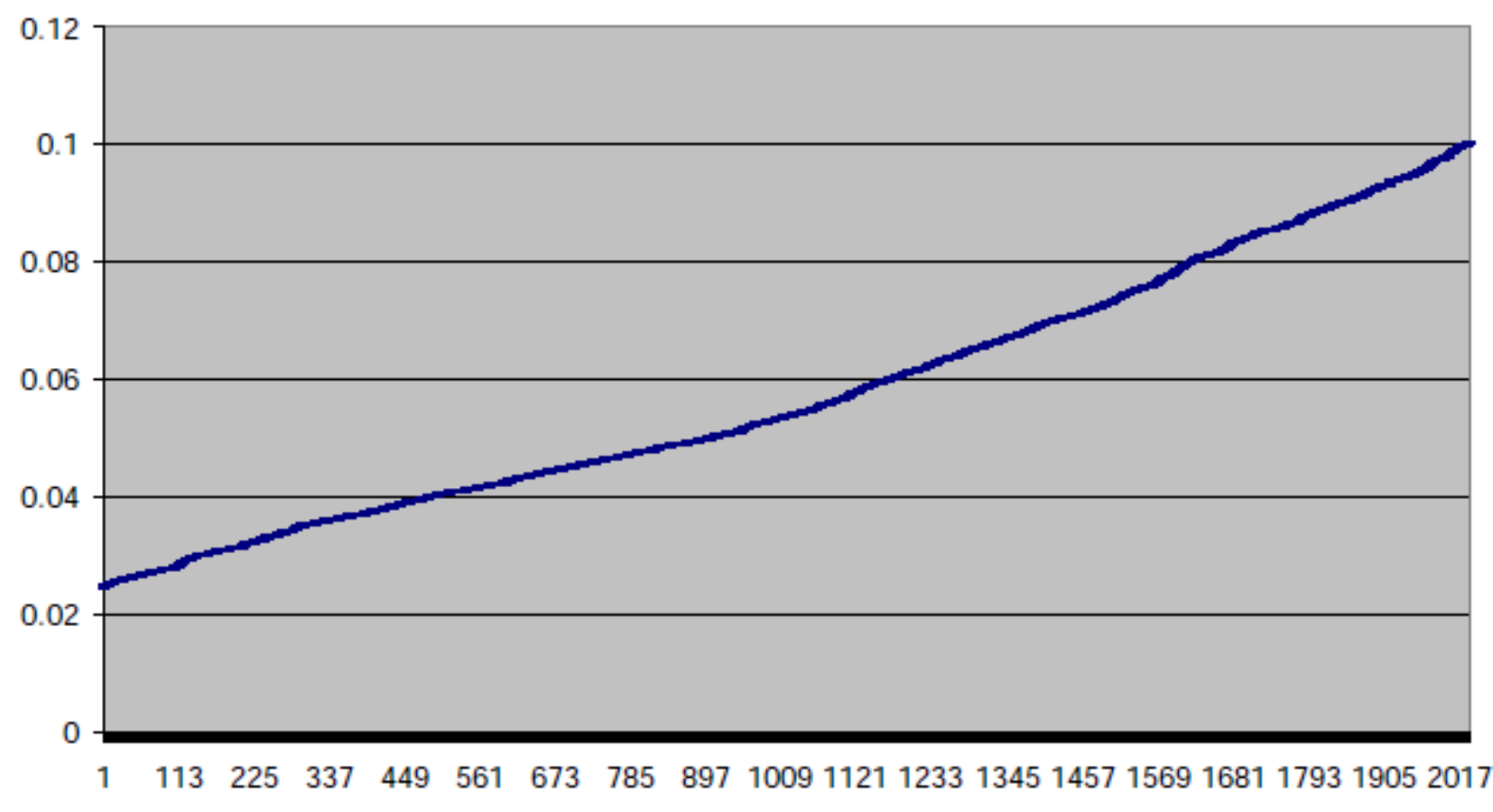}}
\caption{Left~: $P$-over-$N$ architecture for one oscillator. Right~: measured TDC resolution as a function of the combination number for the two oscillators involved in the TDC.}
\label{fig:adj2}
\end{figure}
The final architecture is displayed in Fig.~\ref{fig:adj2} (left scheme). It is a $P$-over-$N$ design for which $C^p_n = \frac{n!}{p!(n-p)!}$ combinations are accessible for each oscillator, and  $\left( C^p_n \right)^2$ for the TDC. Following the calibration procedures described in the previous sections, the TDC resolution has been measured for all possible combinations of a given implementation (Fig.~\ref{fig:adj2}, right). A wide range of resolution is obtained from ~100ps down to ~20ps, with the advantadge of reproducibility which is ensured by the implementation techniques described in the previous sections. This adjustable TDC technique is being published (Girerd 2013).

\subsection{Conclusions}

As demonstrated in muon tomography data analysis applied to volcanology, a good timing resolution is mandatory for background rejection and signal-to-noise ratio improvement. For instance the rejection of the upward-going muons whose trajectories might be confounded with those of downward-going muons crossing the volcano is absolutely necessary to measure the density of the structure under study. This background rejection is based on a time-of-flight analysis. Since the typical size of the instrument is of the order of 1 meter, the time-of-flight of a relativistic muon amounts to some nanoseconds. This imposes severe constraints to the riming resolution of the detector which has limited ressources on field in terms of available power and embedded electronics.\\
In this article we describe the implementation of fine-resolution ring oscillators TDC inside existing FPGA. A simple vernier technique between two oscillators with close frequencies leads to timing resolution down to a tens of pico-seconds. The generic DAQ and clock distribution systems are also detailed. The implementation of the oscillators and the freezing of their routing before the general synthesis is outlined.\\
The implementation on the instrument was performed remotely, while the instrument was on site. We show that simple calibrations may be achieved on site, with the instrument running, without connecting any specific test setup. The use of a common reference clock, already available for synchronization purposes, allows to measure separately one of the two frequencies and the vernier step (i.e. the TDC timing resolution).\\
The technique has been validated on the slopes of La Soufri\`ere de Guadeloupe in the Lesser Antilles and a dedicated commissioning run has been organized to check the presence of upward-going muons by time-of-flight analysis. The results obtained, published separately and summarized in the present article, clearly show the presence of this upward-going muons flux and the rejection power of the TDC data. Direct implications on the quality of the radiographies of the volcano and their consequences on the density analysis are detailed in Jourde et al. (2013).\\

\textbf{Acknowledgments}
Field operations in Guadeloupe received the help from colleagues of the Volcano Observatory, from the crews of the helicopter station of the French Civil Security (\url{www.helicodragon.com}) and from members of the National Natural Park of Guadeloupe (\url{www.guadeloupe-parcnational.fr}). On-field maintenance and servicing of the telescope are ensured by Fabrice Dufour. Field operations on Mount Etna received the help of colleagues of the Volcano Observatory at Catania. Logistic organization was ensured by the \textsc{ulisse-in2p3} department of \textsc{cnrs} (\url{ulisse.cnrs.fr}). We acknowledge the financial support from the UnivEarthS Labex program of Sorbonne Paris Cit\'e (\textsc{anr-10-labx-0023} and \textsc{anr-11-idex-0005-02}). This is IPGP contribution ****.\\

\end{document}